\documentclass[12pt]{article}
\newlength{\dinwidth}
\newlength{\dinmargin}
\usepackage[intlimits,tbtags]{amsmath}
\usepackage{amsfonts}

\newcommand{\be}{\begin{equation}}
\newcommand{\ee}{\end{equation}}
\newcommand{\ber}{\begin{eqnarray}}
\newcommand{\eer}{\end{eqnarray}}

\newcommand{\lp}{\left(}
\newcommand{\rp}{\right)}

\newcommand{\lc}{\left[}
\newcommand{\rc}{\right]}
\newcommand{\dif}{\mathrm{d}}
\newcommand{\2}{\,2}
\newcommand{\bh}{\beta_{\scriptscriptstyle{H}}}
\long\def\symbolfootnote[#1]#2{\begingroup%
\def\thefootnote{\fnsymbol{footnote}}\footnote[#1]{#2}\endgroup}

\begin{document}

\numberwithin{equation}{section}                            
\thispagestyle{empty}

\begin{flushright}
\begin{tabular}{l}
  FFUOV-04/23\\
\end{tabular}
\end{flushright}

\begin{center}
  {\Large \textbf{Exotic Fluids Made of Open Strings}}
\end{center}
\vskip30pt

\centerline{Manuel A. Cobas, M.A.R. Osorio, Mar{\'{\i}}a
Su\'arez
\symbolfootnote[2]{E-mail addresses: cobas, osorio,
maria@string1.ciencias.uniovi.es}}

\vskip6pt 
\centerline{\textit{Dpto. de F{\'{\i}}sica, Universidad de
Oviedo}} 
\centerline{\textit{Avda. Calvo Sotelo 18}}
\centerline{\textit{E-33007 Oviedo, Asturias, Spain}}

\vskip.5in
\begin{center}
  \textbf{Abstract}
\end{center}

We compute the high energy entropy and the equation of state of a gas
of open superstrings in the infinite volume limit focusing on  the
calculation of the number of strings
as a function of energy and volume. We do it in the fixed temperature and fixed
energy pictures to explicitly proof their equivalence. We find that, at high
energy, an effective two dimensional behavior appears for the number of
strings. Looking at the 
equation of
state from a ten dimensional point of view, we show that the Hagedorn behavior
can be seen as correcting the Zeldovich equation of state ($\rho=p$) that can
be found from the two dimensional part of the entropy of the system.
By the way, we show that, near the Hagedorn temperature, the equilibrium state
obtained by sharing  the total energy among
open (super)strings of different length is stable.

\newpage

\section{Introduction}

String theory (with its extensions) hopefully will provide the primordial
matter for the Universe in such a way that, after evolving from a probably non
singular state at very high energy, we find a universe with the
characteristics of the one we observe today. However, almost every important
question is still waiting for a solution from String Theory. Meanwhile, it has
been very clear from the very beginning that String Theory could provide us
with a form of  matter special enough so as to settle what seems to be the main
question which is the one about the resolution of  the initial singularity.

It seems that, to achieve this, we need a kind of reduction of degrees of
freedom at high energy, at least with respect to the contribution coming from
the ultraviolet (high momentum) region. When
only perturbative strings were known, thermal duality showed itself as a
mechanism to truncate ten dimensional string theories (Bosonic,
Heterotic, Type II) at high energy to a kind of effective two dimensional
(massless) field theories with polarizations given by the cosmological
constant of a ten dimensional string theory \cite{atickyotros, miguel}. The
problem was that only for Heterotic Superstrings a finite, but vanishing, high
temperature free energy was found. For other string gases, like the ones made
of Type II Closed Superstrings, thermal duality implies a two dimensional
divergent free energy because the two-dimensional polarization degrees of
freedom are measured by the cosmological constant of the ten dimensional Type-0
Bosonic String. In other cases, like heterotic compactifications, the dual
phase at high energy is non sense since it has positive free energy. Only in
two dimensions, for generic heterotic non supersymmetric strings, we know
of a  meaningful high temperature dual phase \cite{miguel}, but we
are already in a two dimensional space-time at low energy (low temperature).
Anyhow, the high temperature dual phase is very peculiar because it has the
cosmological constant of the theory as the coefficient of $\beta^{\,-2}$ and
still might serve as a high temperature candidate to solve the Hagedorn
behavior through a phase transition.

What we are going to show here is that the gas of open superstrings in a big
nine dimensional container presents, at high energy, a two dimensional behavior
that is corrected by the characteristic dominant Hagedorn 
comportment to produce 
an exotic equation of state of the form $p = \rho/(1 + k \sqrt{\rho})$ that can
finally be approximated by the better known  
$p\propto\sqrt{\rho}$ equation of state.
 The entropic
fundamental relation is given as the sum of the entropy of a gas of massless
particles in two dimensions plus $\beta_H E$ which is the standard
dominant  Hagedorn behavior at high energy for the entropy of any gas made 
of strings (open or closed). The two dimensional contribution to the entropy
alone would produce a Zeldovich's  fluid (also known as stiff fluid) equation
of state $\rho=p$ that has appeared many times, for example in
\cite{BanksFischler}, as a good candidate to be the primordial cosmological
fluid. 

The two dimensional behavior of the system is stressed by the fact that
the number of strings at high energy is the same function of $E$ as in a
massless two dimensional ordinary gas.

\section{The gas of open (super)strings}

It is well known that, perturbatively, the critical behavior of a gas of free
open strings at the Hagedorn temperature is such that the Helmholtz free
energy diverges when approaching $T_H$ from below. In macrocanonical terms,
this means that $T_H$ is a maximum temperature in the sense that any open
string gas in equilibrium that, at vanishing chemical potential, has a
finite internal energy $U(T, V)$ is necessarily kept at a temperature below
$T_H$. On the other hand, in a description at given energy and null chemical
potential, it is very natural to think about a maximum temperature because it
would be a maximum for the temperature as a mathematical function of energy
and volume.

\subsection{The macrocanonical description of the gas of open superstrings}

The black body of free open superstrings in the macrocanonical 
ensemble description has a free energy, $-PV=F\lp \beta,V\rp$,
given by (see, e.g., \cite{emar1})

\be
F\lp \beta\rp =
-\frac{V}{8 \lp 2\alpha'\rp^5}
\int_0^{+\infty}\,\dif t\,t^{-6}\, \lc \theta_3\lp 0,
\frac{\mathrm{i} \beta^2}{4 \alpha' t}\rp -\theta_4 \lp 0,
\frac{\mathrm{i} \beta^2}{4 \alpha' t}\rp \rc\, 
\theta_4 ^{-8}\lp 0, \mathrm{i} t/\pi^2\rp\,\,.
\ee  

Let us remind some things to the reader. 
The exponential growth of the number of states with the mass is here
expressed thorough the behavior of $\theta_4 ^{-8}\lp 0, \mathrm{i}
t/\pi^2\rp$ that certainly diverges when $t$ approaches zero from the right.
The behavior of the fourth Jacobi theta function is encoded in the relation
$\theta_4\lp 0, \mathrm{i}\,/t\rp = t^{1/2}\,\theta_2\lp 0,
\mathrm{i}\,t\rp$ that can be obtained by Poisson resummation of the series
representing the Jacobi theta function.

From all this, one gets that, by using an ultraviolet cutoff,
it is convenient to write the free energy as

\be
\label{separate}
\begin{split}
F\lp \beta\rp \approx & -\frac{V}{8 \lp 2\alpha'\rp^5}
\int_{\epsilon}^{+\infty}\,\dif t\,t^{-6}\, \lc \theta_3\lp 0,
\frac{\mathrm{i} \beta^2}{4 \alpha' t}\rp -\theta_4 \lp 0,
\frac{\mathrm{i} \beta^2}{4 \alpha' t}\rp \rc\,
\theta_4 ^{-8}\lp 0, \mathrm{i} t/\pi^2\rp \\
& -\frac{V}{2^{\,9} \pi^{\,8}\lp 2\alpha'\rp^5}\int_0^{\epsilon}\,\dif
t\,t^{-2} \mathrm{e}^{\lp 8\alpha'\pi^3 - \beta^2\pi\rp/\lp 4\alpha't\rp}\,\,
, 
\end{split}
\ee
where $\epsilon$ is a dimensionless cutoff that is taken small enough  so 
that the second term on the right hand side of eq. \eqref{separate} is a
good approximation to the contribution to the free energy coming
from  the ultraviolet degrees of freedom\footnote{In a quantum field theory,
finite temperature provides an ultraviolet exponential regulator. In String
Theory at finite temperature as described by the analog model, the Hagedorn
behavior appears as an ultraviolet divergence. In the analog model, the
infrared and ultraviolet regions are defined for every quantum field in
the infinite collection of them as vibrational modes of the string. Open
string T-duality exchanges both regimes, but with different objects: a short
distance between D-branes corresponds to a long distance propagation of open
strings.}. Only the Maxwell-Boltzmann contribution has survived, so we are in
the classical statistics approximation. This cut-off can be seen as separating
the infrared from the ultraviolet degrees of freedom. We are going to be interested
on the contribution to the high temperature free energy that precisely comes
from the ultraviolet degrees of freedom (see \cite{atickyotros}).

From here, it is very easy to read the Hagedorn temperature 
for open superstrings (treated without a gauge group contribution)
as $1/\bh=T_H =1/ (\pi \sqrt{8\alpha'})$   

The integral in the second term in eq. \eqref{separate} can be evaluated
to give \cite{97uno}

\be
\int_0^{\epsilon}\,\dif t\,t^{-2}
\mathrm{e}^{\lp 8\alpha'\pi^3 - \beta^{\,2} \pi\rp/\lp 4\alpha't\rp} =
\frac{\bh^{\,2}}{2\pi^3\lp \beta^{\,2}-\bh^{\,2}\rp}\,
\mathrm{e}^{-2\pi^3\lp
\beta^{\,2}-\bh^{\,2}\rp/(\epsilon\bh^{\,2})}\,\, ,
\label{around}
\ee
that is valid when $\beta > \bh$. It is now very clear that the free energy
diverges as $\beta$ approaches $\bh$ and eq. \eqref{around} tells us
exactly how. For a fixed cut-off $\epsilon$, the contribution to the free 
energy coming from the ultraviolet region exponentially falls off when $(\beta
- \bh)$ grows. If we then make the approximation $\lp \beta^2
-\bh^2 \rp = \lp \beta -\bh\rp^2 +2\bh\lp \beta -\bh\rp \approx
2\bh\lp\beta -\bh\rp$, the contribution to $\beta P V = -\beta F\lp
\beta, V\rp = \overline{N}\lp \beta, V\rp$ from the ultraviolet part can be
approximated by\footnote{That $-\beta F\lp \beta, V\rp = \overline{N}\lp \beta,
V\rp$ is the result of implementing $\mu=0$ in the macrocanonical description
when Maxwell-Boltzmann statistics is a good approximation \cite{exten}.}

\be
\begin{split}
-\beta F^{\,h}\lp\beta\rp = &\, \frac{V}{2\pi
\bh^{\,8}}\,\int_0^{+\infty}\,\dif E\,\mathrm{e}^{-\beta E}\,\mathrm{e}^{\bh E}
\theta \lp E - \frac{4 \pi^3}{\epsilon\bh}\rp\\
 = &\,\frac{V}{2\pi \bh^{\,8}}\,\frac{\mathrm{e}^{-\lp\beta
-\bh\rp\,4\pi^{\,3}/\lp\epsilon\bh\rp}}{\beta -\bh}\,\, .
\end{split} 
\label{laplace1} 
\ee

This integral representation tells us several things. First of all it is easy
to read the high energy behavior of the single open string density of
states given by (see \cite{97uno, ska})

\be
\Omega_1^{\,h}\lp V, E\rp =
\frac{V}{2\pi\bh^{\,8}}\,\,\,\mathrm{e}^{\bh E}\,\,\theta\lp E
-\frac{4\pi^3}{\epsilon\bh}\rp , 
\label{omega1}
\ee
where the energy cutoff has been directly derived from the dimensionless
$\epsilon$ parameter.

It is also very clear that the contribution to the free energy near $\beta_H$ is
dominated by $F^{\,h}$ that grows unbounded as long as the first integral in
\eqref{separate}
gives a finite value at $\beta_H$.

The internal energy $U \lp \beta, V\rp$ can easily be computed as 
$U\lp \beta, V\rp = \frac{\partial}{\partial\,\beta} \beta F\lp \beta\rp$.
Near the Hagedorn temperature, we have that

\be
U^{\,h}\lp \beta, V\rp \approx \frac{V}{2\pi \bh^{\,8}}\,
\lc \frac{1}{\lp \beta -\bh\rp^2} - 
\frac{8\pi^{\,6}}{\epsilon^{\,2}\bh^{\,2}}\rc,  
\ee
where we have presented it to show the only divergent term. Let us notice that
$0 < \lp \beta -\bh\rp < \epsilon\bh/\lp \sqrt{8}\pi^{\,3}\rp$.

The calculation of fluctuations is an important piece in the study of
equivalence between different ensemble descriptions. Fluctuations
computed for the energy $U$ are given around $\bh$ by

\be
\frac{\lc T^{\2}\,C_V(T, V)\rc^{1/2}}{U} \approx 2\,\sqrt{\frac{\pi}{V}}
\bh^{\,4}\lp \beta -\bh\rp^{1/2}\,\mathrm{e}^{\lp \beta -\bh\rp
2\pi^3/\lp\epsilon\bh\rp} = \sqrt{2} \lp\overline{N}^{\,h}\rp^{\,-1/2}
, 
\ee 
that certainly vanishes as $\beta$ approaches $\bh$. In fact, fluctuations are
O$(\overline{N}^{-1/2})$ as in a typical closed isothermal system. This
means that equivalence between the fixed temperature and the fixed energy
descriptions for the system of open strings with $\mu=0$ must occur in the
sense that the states  at given energy certainly would correspond to states
with well defined averaged energy at a given temperature.

To compare with the fixed energy (or micro) calculation, it is useful to get
the  thermodynamical fundamental relation giving the entropy $S$ as a function
of the canonical averaged energy $U$ and the volume $V$ for 
$\beta$ bigger and near $\bh$. One gets

\be
S^{\,h}\lp U, V\rp = 2\sqrt{\frac{U\,V}{2\pi\bh^{\,8}}}
+ \bh\,U ,
\label{entropyM}
\ee
where we have assumed that $U \gg 8\,\pi^5\,V/\lp \epsilon^{\,2}\bh^{\,10}\rp$.

Finally, it is worth to remark that the first term in  \eqref{separate}
gives the contribution from the string modes for long propagation times. The
massive modes are exponentially suppressed so the relevant contribution comes
from the massless degrees of freedom \cite{kipfrantz}.

\subsection{The fixed energy description of the gas of open
superstrings}

The big number of open dimensions tells us that 
quantum statistical corrections are negligible for the system \cite{exten}.
Then, the label $N$ in $\Omega_N$ certainly refers to the number of strings in
the gas because Maxwell-Boltzmann statistics is applicable even when the
system energy is not very high.

\begin{figure}[htp]
\let\picnaturalsize=N
\def\picsize{3.0in}
\def\picfilename{abiss.eps}
\ifx\nopictures Y\else{\ifx\epsfloaded Y\else\input epsf \fi
\let\epsfloaded=Y 
\centerline{\ifx\picnaturalsize N
\epsfxsize \picsize\fi\epsfbox{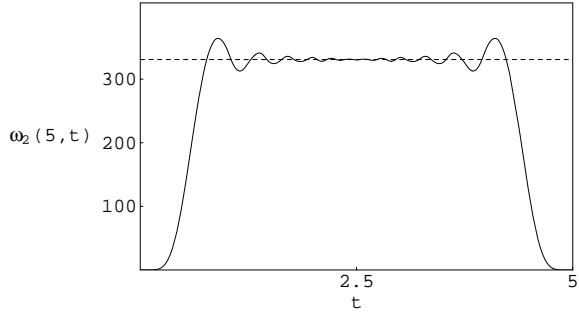}}}\fi
\caption{$\omega_2(5,t)$ computed numerically. The straight line represents
$\omega_2$ as given by using $\Omega_1^h$ only. $\alpha' = 1$.}  
\label{F1}
\end{figure}

$\Omega_1 \lp E, V\rp$ is the key ingredient to get $\Omega_N \lp E, V\rp$
 using Laplace convolutions \cite{97uno}. 
The problem here is that such integrals cannot be performed 
analytically. However, we are interested in getting the high energy
behavior of $\Omega_N$ (for high N). For open strings, this can be
obtained from the convolution of the high energy part of $\Omega_n$ to get
$\Omega_{2n}^{\,h}$, \footnote{In fact,
$\Omega_{2n}\lp E\rp =\frac{n!^{\,2}}{(2n)!}\,\int_{0}^{E}\,\dif\, t \,\,
\Omega_n\lp E - t\rp\,\Omega_n\lp t\rp$ \cite{97uno}.}.
 The reason is that the pure exponential
growth of $\Omega_1^{\,h}$ implies that the convolution between the high and 
low energy parts of the single density of states gives a negligible
contribution to $\Omega_2^{\,h}$ (this is not the case for closed strings when
windings are absent \cite{exten}).
In Fig. \ref{F1}, a numerical computation of $\omega_2
(E, t) = \Omega_1 \lp E-t\rp\,\Omega_1\lp t\rp$ at $E=5$ ($\alpha'=1$) is
showed together with the straight line which represents $\omega_2^{\,h}$ as
obtained  from $\Omega_1^{\,h}\lp E, V\rp \sim \mathrm{e}^{\bh E}$. In it, we
can observe the two energy cutoffs for the validity of this approximation
that, in general, will be placed, for $\omega_2$, at $\lambda$ and $E-\lambda$;
$\lambda = 4\pi^3/\lp\epsilon \bh\rp$. The plateau in the 
interval $[\lambda,5-\lambda]$ signals the feature that 
energy will be distributed with equal probability among open strings of 
different length, i.e., different energy. This is the
 way the equilibrium state
shows the most probable decay of a highly excited open string 
(see \cite{manes}).  Looking at this plateau from the point of view
of equipartition and the equivalence of ensembles, one might erroneously guess
that the fixed temperature and preserved energy descriptions would not be
equivalent. Instead, one would expect as the most probable configuration that
each open string
shared half of the total energy. However, the computation of energy
fluctuations tells us that this is not the case. 
We are actually showing that the equilibrium got by shearing
the total energy among open strings with different lengths is stable. Indeed
this is a worthy feature of the behavior, around the Hagedorn temperature, of
our system.

With this approximation one can get 

\be
\begin{split}
\Omega_N^{\,h} \lp E, V\rp = &\frac{1}{N!\,\lp N-1\rp!}\lp \frac{V}{2\pi
\bh^{\,8}}\rp^{\,N}\mathrm{e}^{\bh E}\lp
E-N\lambda\rp^{\,N-1}\,\theta\lp E-N\lambda\rp\\
\approx &
\frac{1}{N!^{\,2}}\lp \frac{V}{2\pi \bh^{\,8}}\rp^{\,N}\mathrm{e}^{\bh E}\lp
E-N\lambda\rp^{\,N}\,\theta\lp E-N\lambda\rp,
\end{split} 
\ee
where in the second equation we have assumed that $N$ is very big. 
This high energy density of states for the gas of $N$ strings holds as long as
$E$, the energy of the gas, is bigger than $N\lambda$.

The vanishing of the chemical potential $\mu$ implies that the high energy
entropy is $S^{\,h}=\mathrm{ln}\sum_{N=0}^{\infty}\, \Omega_N^{\,h}$. In fact,
the sum can be approximated by a single term $\Omega_{N^{\,*}}^{\,h}$ for
$N^{\,*}$ such that $\Omega_{N=N^{\,*}}$ is a maximum of the density of
states  as a function of the number of particles. We then compute
$\frac{\partial}{\partial N}\,\mathrm{ln}\Omega_N^{\,h}$ to get that, for $E
\gg \lambda N^{\,*}$, 

\be
 \overline{N}^{\,h}\lp E, V\rp= N^{\,*}= \sqrt{\frac{E\,V}{2\pi\bh^{\,8}}}.
 \label{ene}
\ee
It is a very notorious fact that the number of open superstrings at high
energy depends on energy and volume as in a  gas of massless particles in two
space-time dimensions described using Maxwell-Boltzmann
statistics\footnote{For the  gas of massless particles in $d > 2$ space-time
dimensions with Bose-Einstein statistics, one finds: $\overline{N}\lp E, V\rp =
\frac{\zeta\lp d-1\rp}{\zeta\lp d\rp}\,N^{\,*}\lp E, V\rp$ where $N^{*}$ is
the number of particles with $\mu=0$ using Maxwell-Boltzmann statistics
\cite{exten}. In two space-time dimensions, one gets the finite result:
$\overline{N}\lp E, L\rp = \frac{\sum_{r=1}^{+\infty}\,\theta\lp ER
-r\rp/r}{\zeta\lp 2\rp}\,N^{\,*}\lp E, L\rp$, where $R$ is the radius of the
compact space ($L=2\pi R$) that, as the energy, is supposed to become big as
one takes the thermodynamic limit. This  expression at dimension two
results from computing the free energy in the thermodynamic limit by
converting the sum over discrete momenta to an integral together with the
separation of the zero momentum contribution that represents the one of the
the vacuum state (no particle state) that must not be second
quantized.}. 

The entropy is then

\be
S^{\,h}\lp E, V\rp \approx 2\,\overline{N}^{\,\,h}\lp E, V\rp + \beta_H\,E = 2
\sqrt{\frac{E\,V}{2\pi\bh^{\,8}}} + \bh E, \label{entropym}
\ee
for $E \gg 8\pi^{\,5}\,V/\lp\epsilon^{\,2}\bh^{\,10}\rp$.
It exactly coincides with $S^{\,h}\lp U, V\rp$ as computed in the canonical
description, as we expected from the computation of the energy fluctuations in
the fixed temperature picture. This expression of the entropy as a function
of energy and volume was found in \cite{9408134}, \cite{9707167} (see also
\cite{9902058}), although no physical interpretation of the term which scales
with $E^{1/2}$ in terms of the
number of strings was given\footnote{The equivalence of ensembles was also not
treated. It would be a mistake to think that it is trivial because the
specific
heat is positive and it would be cynic to say that equivalence  has been
implicitly
assumed since
the computation of energy fluctuations is so easy that does not deserve
any mention. After all, we go further because we show here that the equilibrium
got shearing the
total energy with equal probability among a set of open strings with different
lengths, i.e. different energies, is compatible with the stability of the
system. This is more
than
stating that equivalence holds, it is a proof of the thermodynamic stability of
the stringy
matter that comes from the preferred decay mode of any highly excited open
string \cite{manes}.}. It is also worth to mention that this entropy
is an extensive function of energy and volume that can be  written in terms
of the number of strings as

\be
S^{\,h} = 2\overline{N} + 2\pi\bh^{\,9}\overline{N}^{\,2}/V
\label{ene}
\ee

This behavior in terms of the number of objects can be compared with a "regular"
system for which entropy is $O(N)$.
In \eqref{entropym} and \eqref{ene}, the first term is very easily identified
as 
the entropy for a two dimensional gas of massless particles that can be 
written as a
function of the squared root of the energy or, exactly, as twice the average
number of objects\footnote{The massless character of
the particles does not seem to be something special because the high
temperature limit for the free energy of a massive field in $d$ space-time
dimensions is dominated by the  contribution to the free energy of a massless
field in $d$ dimensions.}. We also
see the nine dimensional volume divided by
the adequate power of the length scale of our problem which is the squared
root of $\alpha'$ (included in $\beta_H$). This is the way an
effective one
dimensional volume appears. This contribution to the entropy is corrected by the
universal volume independent, and then pressureless, contribution coming from
the Hagedorn behavior that is $O(\overline{N}^{\,2})$. It is important to
notice that the term $\beta_H\,E$
shows up in the entropic fundamental relation for any critical string gas in the
macrocanonical and the microcanonical computations. This term gives, for any
gas of fundamental strings, a divergent contribution to $C_V(E)$ and the
effective two dimensional part finally makes $C_V(E)$ finite and positive for
the gas of open superstrings. The two terms in the entropy can be seen as the
contribution of two different types of degrees of freedom at high energy that
finally make $T_H$ a maximum temperature for the system.

\section{The equation of state}

It is very easy to get the equation of state which relates the density of
energy $\rho$ and the pressure of the gas. We get

\be
P = \frac{\rho}{1\,+\,\beta_H\,\sqrt{2\pi\beta_H^{\,8}\,\rho}}\,\, .
\ee
Here, what makes the denominator different from the unity, and so the equation 
of state different from Zeldovich's one, is precisely the term $\beta_H\,E$
in the entropy. This is dominant at high energy, because what we mean by high
energy to get \eqref{entropyM} and \eqref{entropym} implies the condition
$\sqrt{2\pi \beta_H^{\,10}\,\rho} \gg 1$.

With this approximation the equation of state at very high energy looks

$$
P \approx \sqrt{\frac{\rho}{2\pi\beta_H^{\,10}}}\,\, .
$$
This fluid is causal because the sound would propagate in it with a speed
given by

$$
v_s^{\,2} = \frac{\partial P}{\partial \rho} \approx \frac{1}{2\,
\sqrt{2\pi\beta_H^{\,10}\rho}}\ll 1.
$$

\section*{Acknowledgments}
This work has been partially supported by the Spanish MEC research project
BFM2003-00313/FISI. The work of M. A. C. is partially supported by a 
MEC-FPI fellowship. The work of M. S. is partially supported by a MEC-FPU
fellowship.

\end{document}